\documentclass[prb,twocolumn,showpacs]{revtex4}
\usepackage{graphicx} 

\usepackage{amsmath,amsthm,amssymb,mathrsfs}
\usepackage{bm}

\begin{document}

\title{Chaos in nanomagnet via feedback current}

\author{Tomohiro Taniguchi${}^{1}$, Nozomi Akashi${}^{2}$, Hirofumi Notsu${}^{3,4}$, Masato Kimura${}^{3}$, Hiroshi Tsukahara${}^{5}$, and Kohei Nakajima${}^{2}$
      }
 \affiliation{
 ${}^{1}$National Institute of Advanced Industrial Science and Technology (AIST), Spintronics Research Center, Tsukuba, Ibaraki 305-8568, Japan, \\
 ${}^{2}$Graduate School of Information Science and Technology, The University of Tokyo, Bunkyo-ku, 113-8656 Tokyo, Japan, \\
 ${}^{3}$Faculty of Mathematics and Physics, Kanazawa University, Kanazawa, Ishikawa 920-1164, Japan, \\
 ${}^{4}$JST, PRESTO, 4-1-8 Honcho, Kawaguchi, Saitama 332-0012, Japan, \\
 ${}^{5}$High Energy Accelerator Research Organization (KEK), Tsukuba, Ibaraki 305-0801, Japan
 }

\date{\today} 
\begin{abstract}
{
Nonlinear magnetization dynamics excited by spin-transfer effect with feedback current is studied both numerically and analytically. 
The numerical simulation of the Landau-Lifshitz-Gilbert equation indicates 
the positive Lyapunov exponent for a certain range of the feedback rate, which identifies the existence of chaos in a nanostructured ferromagnet. 
Transient behavior from chaotic to steady oscillation is also observed in another range of the feedback parameter. 
An analytical theory is also developed, 
which indicates the appearance of multiple attractors in a phase space due to the feedback current. 
An instantaneous imbalance between the spin-transfer torque and damping torque causes a transition between the attractors, and results in the complex dynamics. 
}
\end{abstract}

 \maketitle


\section{Introduction}
\label{sec:Introduction}

Nonlinear dynamics can be found in a wide variety of physical, chemical, and biological systems from small to large scale [\onlinecite{strogatz01,pikovsky03}]. 
Recent observations of rich magnetization dynamics, such as switching, auto-oscillation (limit cycle), and synchronization, excited in a nanostructured ferromagnet 
have also proved the applicability 
of nonlinear science to a fine structure [\onlinecite{katine00,kiselev03,rippard04,bertotti05,slavin05PRL,houssameddine07,kim08a,kubota13,grimaldi14,awad17}]. 
These dynamics are driven by spin current carried by, for example, conducting electrons in metals [\onlinecite{slonczewski96,berger96,ralph08}]. 
Since the spin current in metals can survive only within nanometer scale [\onlinecite{valet93}], 
these magnetization dynamics had not been observed until the development of fabrication technology of nanostructure was achieved. 
A new direction investigating the applicability of such nonlinear magnetization dynamics to 
non-von Neumann computing scheme, inspired by human brain, has been growing very recently [\onlinecite{grollier16,torrejon17,kudo17,furuta18}]. 


An attractive and intriguing phenomenon in nonlinear science is chaos [\onlinecite{alligood97,ott02}]. 
It should be noticed here that the previous works in magnetism and spintronics have clarified that the magnetization dynamics in a nanostructured ferromagnet 
is sufficiently sufficiently well described by two dynamical variables [\onlinecite{sun00,grollier03,tatara04,guslienko06,slavin09,bertotti09text}]. 
For example, the macrospin model has two dynamical variables describing the zenith and azimuth angles of the magnetization. 
The Thiele equation depicting the magnetic vortex or skyrmion dynamics includes two variables corresponding to the radius and phase of the core, 
whereas the domain wall motion is represented by the center of the wall position and the tilted angle of the magnetization at the center. 
On the other hand, according to the Poincar\'e-Bendixson theorem, chaos is prohibited in a two-dimensional system [\onlinecite{alligood97}]. 
Therefore, an additional degree of freedom is necessary to induce chaos in ferromagnets. 
In previous works, chaos has been studied for systems with alternative current [\onlinecite{li06,yang07}] or 
magnetic and/or electric interaction between two ferromagnets [\onlinecite{kudo06,taniguchi19}]. 
The former makes the system nonautonomous, whereas the latter utilizes many-body system. 
Another possible source causing highly nonlinear dynamics is feedback force with delay 
because the presence of the delay makes the dimension of the system infinite [\onlinecite{mackey77}]. 
Recently, the modulation of the threshold current by the self-injection of the feedback current into the vortex ferromagnet was 
theoretically predicted [\onlinecite{khalsa15}] and was experimentally confirmed [\onlinecite{tsunegi16}]. 
Complex dynamics in an in-plane magnetized ferromagnet with feedback current was also found by numerical simulation [\onlinecite{williame19}]. 
However, it should be emphasized that the existence of the feedback effect does not necessarily guarantee chaos. 
Therefore, a careful analysis is necessary for the magnetization dynamics in the presence of feedback effect in order to identify chaos. 


The purpose of this work is to develop a theoretical analysis of the nonlinear magnetization dynamics in a nanostructured ferromagnet in the presence of feedback current. 
We perform the numerical simulation of the Landau-Lifshitz-Gilbert (LLG) equation in spin torque oscillator (STO), 
and find that the feedback current causes highly nonlinear dynamics of the magnetization. 
This work identifies chaos by the positive Lyapunov exponent, which is found in a certain range of the feedback rate, 
whereas transient behavior is also observed in another range of the feedback rate. 
We also develop an analytical theory to reveal the origin of such complex dynamics. 
The bifurcation analysis indicates that the feedback current results in the appearance of multiple attractors in the phase space. 
An instantaneous imbalance between the spin-transfer torque and damping torque allows a transition between these attractors, 
and induces the complex dynamics found in the numerical analysis. 


The paper is organized as follow. 
In Sec. \ref{sec:System description}, we describe the structure of the STO and show the LLG equation including feedback current. 
In Sec. \ref{sec:Numerical analysis}, the results of the numerical simulation of the LLG equation are presented. 
The Lyapunov exponents and bifurcation diagrams as functions of the feedback rate and delay time are also presented. 
In Sec. \ref{sec:Theoretical analysis}, a theoretical analysis on a multiple attractor is discussed. 
Section \ref{sec:Conclusion} summarizes this work.


\section{System description}
\label{sec:System description}

In this section, we describe the system under consideration, 
and provide the comment on the numerical methods. 
The details of the algorithms are also given in the Supplemental Material [\onlinecite{comment_suppl}] (which includes Ref. [\onlinecite{kanno14}]). 



\begin{figure}
\centerline{\includegraphics[width=1.0\columnwidth]{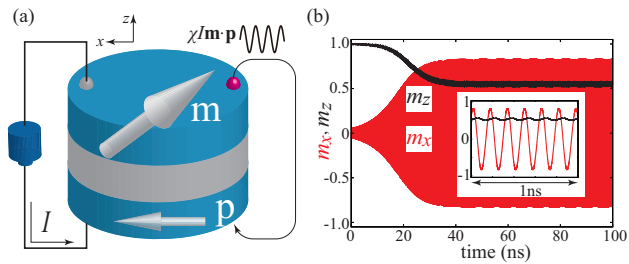}}
\caption{
        (a) Schematic view of the system. 
            The direct current $I$ is injected from the reference layer to the positive layer, 
            whereas the current, $\chi I \mathbf{m}\cdot\mathbf{p}$, outputted from the STO is injected into the STO with time delay $\tau$. 
            The feedback current oscillates when the magnetization $\mathbf{m}$ in the free layer is in a dynamical state. 
        (b) Typical magnetization dynamics in the absence of the feedback current. 
            The inset shows an auto-oscillation in a steady state. 
         \vspace{-3ex}}
\label{fig:fig1}
\end{figure}



\subsection{LLG equation}
\label{sec:LLG equation}

The system under consideration is schematically shown in Fig. \ref{fig:fig1}(a). 
The unit vectors pointing in the magnetization directions in free and reference layers are denoted as $\mathbf{m}$ and $\mathbf{p}$, respectively. 
Direct current, $I$, is injected from the reference to free layer, 
and excites an auto-oscillation of the magnetization $\mathbf{m}$ via spin-transfer effect [\onlinecite{slonczewski96,berger96}]. 
Here, we focus on the STO consisting of a perpendicularly magnetized free layer and an in-plane magnetized reference layer 
because this type of STO can emit large emission power with narrow linewidth [\onlinecite{kubota13}], 
and therefore, is of great interest from viewpoints of both fundamental and applied physics. 
The magnetization $\mathbf{p}$ in the reference layer points to the positive $x$ direction, 
whereas the $z$ axis is perpendicular to the film-plane. 
The magnetization dynamics in the free layer is described by the Landau-Lifshitz-Gilbert (LLG) equation given by 
\begin{equation}
  \frac{d \mathbf{m}}{dt}
  =
  -\gamma
  \mathbf{m}
  \times
  \mathbf{H}
  -
  \gamma
  H_{\rm s}
  \mathbf{m}
  \times
  \left(
    \mathbf{p}
    \times
    \mathbf{m}
  \right)
  +
  \alpha
  \mathbf{m}
  \times
  \frac{d\mathbf{m}}{dt}
  \label{eq:LLG}
\end{equation}
where $\gamma$ and $\alpha$ are the gyromagnetic ratio and the Gilbert damping constant respectively. 
The magnetic field $\mathbf{H}=[H_{\rm appl}+(H_{\rm K}-4\pi M)m_{z}]\mathbf{e}_{z}$ consists of 
an applied field $H_{\rm appl}$, interfacial magnetic anisotropy field $H_{\rm K}$ [\onlinecite{yakata09,ikeda10,kubota12}], and demagnetization field $-4\pi M$. 
The spin-transfer torque strength, $H_{\rm s}$ is given by 
\begin{equation}
  H_{\rm s}
  =
  \frac{\hbar \eta I [ 1 + \chi \mathbf{m}(t-\tau)\cdot\mathbf{p}]}{2e(1+\lambda\mathbf{m}\cdot\mathbf{p})MV}, 
  \label{eq:H_s}
\end{equation}
where $M$ and $V$ are the saturation magnetization and the volume of the free layer, respectively. 
The spin-transfer torque strength is characterized by the spin polarization $\eta$ and spin-transfer torque asymmetry $\lambda$. 
The values of the parameters used in this work are derived from the experiment [\onlinecite{kubota13}], as well as a theoretical analysis [\onlinecite{taniguchi17}] as 
$M=1448.3$ emu/c.c., $H_{\rm K}=18.616$ kOe, $H_{\rm appl}=2.0$ kOe, $V=\pi \times 60^{2} \times 2$ nm${}^{3}$, 
$\eta=0.537$, $\lambda=0.288$, $\gamma = 1.764 \times 10^{7}$ rad/(Oe s), and $\alpha=0.005$. 
The current of $I=1.0$ mA corresponds to the current density of $8.8$ MA/cm${}^{2}$. 
An auto-oscillation in the absence of the feedback is excited in this type of STO 
when the current magnitude becomes larger than a threshold value [\onlinecite{taniguchi17}] 
(see also Appendix \ref{sec:AppendixA} for derivation), 
\begin{equation}
  I_{\rm c}
  =
  \frac{4\alpha eMV}{\hbar \eta \lambda}
  \left(
    H_{\rm appl}
    +
    H_{\rm K}
    -
    4\pi M
  \right), 
  \label{eq:critical_current}
\end{equation}
which is about $1.6$ mA for the present parameters. 
Figure \ref{fig:fig1}(b) shows a typical magnetization dynamics in the absence of the feedback current, where the direct current is $I=2.5$ mA. 
As shown, an auto-oscillation having a period of $0.16$ ns is excited after a relaxation time on the order of $10$ ns. 
The inset of Fig. \ref{fig:fig1}(b) shows the dynamics of $m_{x}$ (red) and $m_{z}$ (black) in a steady state. 
It can be seen from the figure that $m_{z}$ is almost temporally constant but slightly oscillates around a certain value. 
These results will be used for comparison with the dynamics in the presence of the feedback current, as well as for the development of an analytical theory, below. 


\subsection{Description of feedback effect}
\label{sec:Description of feedback effect}

The strength of the spin-transfer torque, Eq. (\ref{eq:H_s}), includes the feedback current given by $\chi I \mathbf{m}(t-\tau)\cdot\mathbf{p}$, 
where $\chi$ is the rate of the feedback current with respect to the direct current $I$, 
whereas $\tau$ is the delay time. 
Due to tunnel magnetoresistance effect, the feedback current depends on the relative direction of the magnetizations, $\mathbf{m}\cdot\mathbf{p}$ [\onlinecite{kubota13}]. 
The feedback current brings the past information of the magnetization state, and extends the dimension of the phase space, 
which presents a possibility to excite chaos in STO. 

Let us give brief comments on experiment to measure chaos in an STO. 
An experimental work injecting the feedback current to a vortex STO and measuring the output power was already reported [\onlinecite{tsunegi16}]. 
The feedback current can be injected to the STO independently from the direct current by using a bias-Tee and delay line. 
The numerical analyses shown below, as well as the analytical theory developed in Sec. \ref{sec:Theoretical analysis}, predict that 
chaos appears for a large feedback rate $\chi$ and/or long delay time $\tau$ compared to typical time scales of the STO. 
The typical value of the delay time possible in experiment is on the order of $10$ ns [\onlinecite{tsunegi16}]. 
On the other hand, the oscillation period ($\sim 3$ ns) of the vortex STO used in the previous work [\onlinecite{tsunegi16}] is only 10 times shorter than the delay time. 
This might be the reason why chaos was not confirmed in the previous works. 
Regarding this point, two approaches are taken into account to observe chaos in STO. 
The first one is to make the delay time long. 
A long delay time is realized by using a long electric cable. 
The second approach is to use an STO having a short oscillation period. 
In fact, the STO studied in this work has a short period because of macrospin structure of the magnetization. 
Therefore, the theoretical analyses shown below will possibly be examined experimentally.
A possible remaining issue, however, may be an energy loss in a cable, which should be optimized in experiments.

We also give a comment on the method to identify chaos by experiments. 
The experimental methods to identify chaos are, for example, 
the estimation of the Lyapunov exponent from time series of data and/or Fourier analysis. 
The former method requires to measure the dynamical trajectory in the system, 
and estimate the Lyapunov exponent from a discrete set of time series data by evaluating the principal axis of the expansion [\onlinecite{wolf85}]. 
A possible problem in applying this method to STO is the limitation of the information on the dynamical trajectory obtained. 
The magnetization dynamics in the STO is measured through the magnetoresistance effect. 
Both giant and tunnel magnetoresistances are proportional to $\mathbf{m}\cdot\mathbf{p}$. 
Therefore, we can measure only the component of the magnetization $\mathbf{m}$ projected to the direction of $\mathbf{p}$. 
This fact might make it difficult to reproduce the dynamical trajectory and identify chaos from the time series of data. 
The Fourier analysis, on the other hand, indicates chaos from the shape of the spectrum. 
The Fourier spectrum shows a sharp peak for a non-chaotic dynamics, whereas it has a broad structure without a unique peak in chaos state; 
see also Sec. \ref{sec:Lyapunov exponent as a function of feedback rate}. 
Therefore, the Fourier spectrum provides an evidence to identify chaos. 



\subsection{Numerical method}
\label{sec:Numerical method}

Here, let us provide a brief description of the numerical technique used in the next section. 
The LLG equation, Eq. (\ref{eq:LLG}), with the feedback current is solved by a fourth-order Runge-Kutta scheme accompanied with continuation method. 
The details of this algorithm are summarized in the Supplemental Material [\onlinecite{comment_suppl}]. 
We also evaluate the bifurcation diagram, which is defined as the local maximum of $m_{z}(t)$ after the magnetization moves to an attractor. 
In this work, chaos is defined as the dynamics with the positive Lyapunov exponent. 
The Lyapunov exponent in this work is defined as an average of the instantaneous expansion rate of the dynamical trajectory in the phase space 
with respect to a small perturbation $\epsilon$ as 
\begin{equation}
  \tilde{\lambda}
  =
  \lim_{N_{\rm L}\to \infty}
  \frac{1}{N_{\rm L} \Delta t}
  \sum_{i=1}^{N_{\rm L}}
  \log
  \bigg|
    \frac{\tilde{\epsilon}_{i}}{\epsilon}
  \bigg|,
  \label{eq:Ly_def}
\end{equation}
where $\Delta t$ is the time step of the LLG simulation. 
The number of the perturbation applied to the STO is $N_{\rm L}$, whereas $\tilde{\epsilon}_{i}$ is the expansion of the dynamical trajectory with respect to the $i$th perturbation. 
The detail of the algorithm to evaluate the Lyapunov exponent is also summarized in the Supplemental Material [\onlinecite{comment_suppl}].



\begin{figure*}
\centerline{\includegraphics[width=2.0\columnwidth]{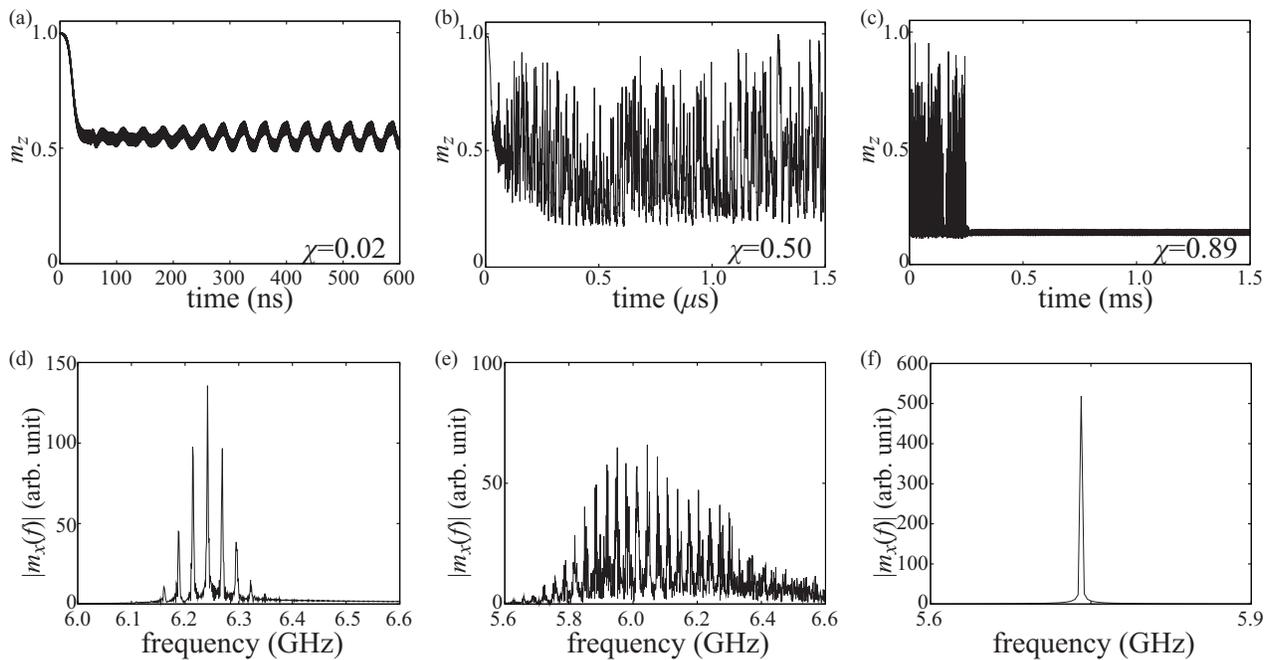}}
\caption{
         Time evolutions of the perpendicular component $m_{z}(t)$ for the feedback rates of (a) $\chi=0.02$, (b) $0.50$, and (c) $0.89$. 
         The current and the delay time are $I=2.5$ mA and $\tau=30$ ns. 
         Note that the time range of each figure is different. 
         Fourier spectra of the in-plane component $m_{x}(t)$ for (d) $\chi=0.02$, (e) $0.50$, and (f) $0.89$ are also shown. 
         \vspace{-3ex}}
\label{fig:fig2}
\end{figure*}




\section{Numerical analysis}
\label{sec:Numerical analysis}

In this section, we show the results of the numerical simulation of the LLG equation, as well as the Lyapunov exponent and bifurcation diagram. 


\subsection{Lyapunov exponent as a function of feedback rate}
\label{sec:Lyapunov exponent as a function of feedback rate}

Here, we show the Lyapunov exponent and the birfurcation diagram as a function of the feedback rate $\chi$. 
The value of $\tau$ in this section is set to be $30$ ns. 
Figures \ref{fig:fig2}(a)-\ref{fig:fig2}(c) show the time evolutions of $m_{z}(t)$ for $\chi=0.02$, $\chi=0.50$, and $\chi=0.89$, respectively. 
Note that the time range of each figure is different to understand the characteristics of each dynamics. 
In the presence of a small feedback current shown in Fig. \ref{fig:fig2}(a), 
although the amplitude of the oscillation is modulated, the dynamics in the steady state is still periodic. 
On the other hand, when the feedback rate becomes relatively large, 
chaotic behavior appears, as shown in Fig. \ref{fig:fig2}(b). 
In this case, non-periodic and highly nonlinear dynamics appears over a wide time range. 
The value of $m_{z}$ oscillates almost over its possible value, $|m_{z}| \le 1$. 
A further increase of the feedback rate leads to a transition of the magnetization dynamics from chaotic to non-chaotic, as shown in Fig. \ref{fig:fig2}(c). 
The chaotic dynamics suddenly disappears 
after a comparatively long period, i.e., longer than the oscillation period of the limit cycle in the absence of the feedback current. 
As mentioned below, the Lyapunov exponents of the dynamics in Figs. \ref{fig:fig2}(a) and \ref{fig:fig2}(c) are zero, 
whereas it is positive for the dynamics in Fig. \ref{fig:fig2}(b). 


Note that evaluating the perpendicular component $m_{z}$ in time domain is useful for theoretical analysis 
because it is approximately constant in the absence of the feedback effect, whereas it becomes complex by the feedback force, as mentioned above. 
On the other hand, evaluating the in-plane component $m_{x}$ will be useful for experiments 
because it directly relates to the experimentally observed signal through magnetoresistance effect. 
Therefore, we also show the Fourier spectra of $m_{x}$ for $\chi=0.02$, $0.50$, and $0.89$ in Figs. \ref{fig:fig2}(d)-\ref{fig:fig2}(f), respectively. 
The Fourier spectrum has a sharp peak with subpeaks for $\chi=0.02$, which is a typical spectrum of the oscillation with the amplitude modulation. 
The Fourier spectrum for $\chi=0.50$, on the other hand, shows a broad structure over a relatively wide range of the frequency. 
A main peak is not uniquely determined. 
The structure implies that the dynamics is chaos. 
The Fourier spectrum for $\chi=0.89$ shows a sharp peak, corresponding to the oscillation frequency after the transition from chaotic to limit cycle oscillation. 
The oscillation frequency is different from that in the absence of the feedback 
because the oscillation amplitude is modified due to the feedback effect. 
Regarding these results, the Fourier analysis will be a possible tool to experimentally identify chaos. 




\begin{figure}
\centerline{\includegraphics[width=1.00\columnwidth]{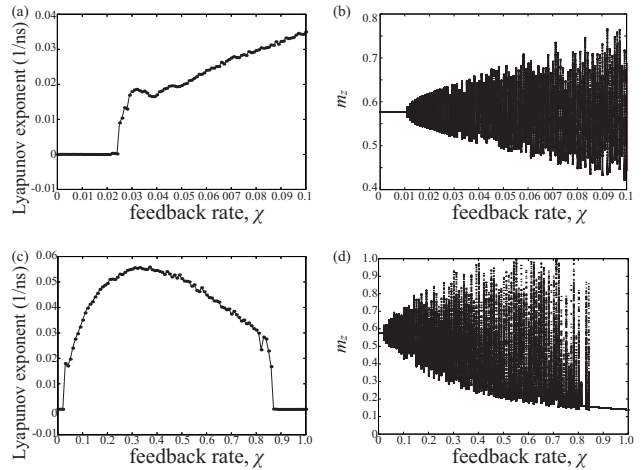}}
\caption{
         (a) (Maximum) Lyapunov exponent and (b) bifurcation cascade (local maximum of $m_{z}$) as a function of the feedback rate $\chi \le 0.10$. 
         The current and delay time are $I=2.5$ mA and $\tau=30$ ns, respectively. 
         The range of $\chi$ is extended to $\chi \le 1.00$ in (c) and (d). 
         \vspace{-3ex}}
\label{fig:fig3}
\end{figure}

Figures \ref{fig:fig3}(a) and \ref{fig:fig3}(b) show the Lyapunov exponent and the bifurcation diagram as a function of the feedback rate in a small range $\chi \le 0.10$. 
The Lyapunov exponent remains zero for $\chi \lesssim 0.024$, where the dynamics is a limit cycle, such as shown in Fig. \ref{fig:fig1}(b), 
or the oscillation with an amplitude modulation as shown in Fig. \ref{fig:fig2}(a). 
In the limit cycle state, the local maximum of $m_{z}$ is a single value, 
whereas it takes several values and shows symmetric distributions around its center in the modulated dynamics, as can be seen in Fig. \ref{fig:fig3}(b). 
The Lyapunov exponent becomes positive for $\chi \gtrsim 0.025$, where the bifurcation diagram shows an inhomogeneous (asymmetric) structure. 
The Lyapunov exponent and the bifurcation diagram for a wide range of the feedback rate, $\chi \le 1.00$, are shown in Figs. \ref{fig:fig3}(c) and \ref{fig:fig3}(d), respectively. 
The positive Lyapunov exponent indicates the existence of chaos in STO. 
The Lyapunov exponent becomes zero again when the feedback rate is further increased to $\chi \simeq 0.87$. 
The magnetization dynamics shown in Fig. \ref{fig:fig2}(c), corresponding to this parameter region, 
can be regarded as transient chaos, which can be found in, for example, a spatially extended turbulence model [\onlinecite{crutchfield88}], 
where the dynamical system finally arrives at an attractor with zero or negative Lyapunov exponent long time after showing chaotic behavior [\onlinecite{alligood97}].
For example, the transient time observed in Fig. \ref{fig:fig2}(c) is on the order of 0.1 ms, 
which is sufficiently longer than the period of the auto-oscillation in the absence of the feedback current (0.16 ns) 
but is measurable because it is shorter than the experimentally available time range for STO dynamics reported up to date, 1.6 ms [\onlinecite{tsunegi18}]. 


\subsection{Lyapunov exponent as a function of delay time}
\label{sec:Lyapunov exponent as a function of delay time}

\begin{figure}
\centerline{\includegraphics[width=1.0\columnwidth]{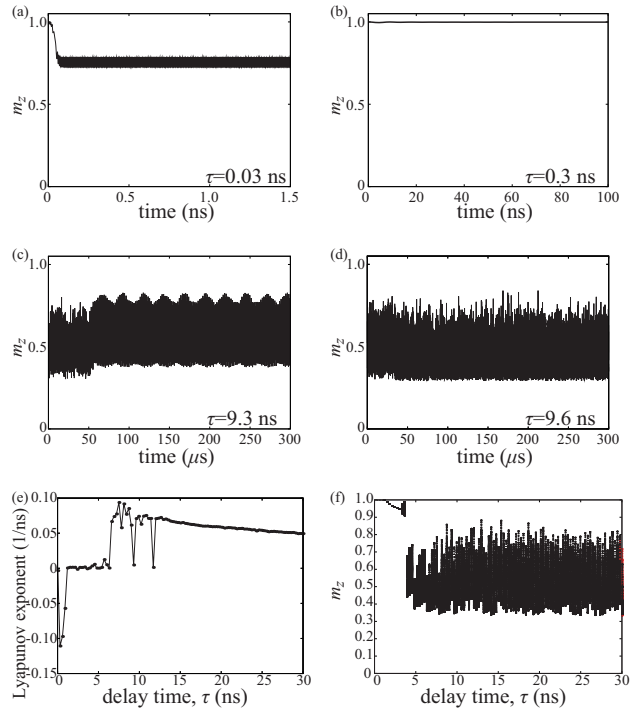}}
\caption{
         Time evolutions of $m_{z}(t)$ for the delay times of (a) $\tau=0.03$, (b) $0.3$, (c) $9.3$ ns, and (d) $9.6$ ns. 
         The current and the feedback rate are $I=2.5$ mA and $\chi=0.20$. 
         (e) The Lyapunov exponent and (f) bifurcation cascade (local maximum of $m_{z}$) as a function of the delay time. 
         \vspace{-1ex}}
\label{fig:fig4}
\end{figure}

Here, we show the Lyapunov exponent and the birfurcation diagram as a function of the delay time $\tau$. 
The value of $\chi$ in this section is set to be $0.20$. 
Figures \ref{fig:fig4}(a) and \ref{fig:fig4}(b) show the time evolutions of $m_{z}$ for short delay times, $\tau=0.03$ and $0.3$ ns, respectively. 
For such a sufficiently short delay time, 
the current necessary to excite an auto-oscillation of the magnetization is given by 
(see also Appendix \ref{sec:AppendixA} for derivation) 
\begin{equation}
  \tilde{I}_{\rm c}
  =
  \frac{4\alpha eMV}{\hbar \eta \lambda p_{0}}
  \left(
    H_{\rm appl}
    +
    H_{\rm K}
    -
    4\pi M
  \right),
  \label{eq:Ic_tilde}
\end{equation}
where $p_{0}=p(\chi,\tau,\theta=0)$ is 
\begin{equation}
  p_{0}
  =
  1
  -
  \frac{\chi}{\lambda}
  \cos 2\pi f_{\rm FMR} \tau,
  \label{eq:p_zero}
\end{equation}
where $f_{\rm FMR}=\gamma(H_{\rm appl}+H_{\rm K}-4\pi M)/(2\pi)$ is the ferromagnetic resonance (FMR) frequency. 
In the absence of the feedback current ($\chi \to 0$), Eq. (\ref{eq:Ic_tilde}) becomes identical to Eq. (\ref{eq:critical_current}). 
According to Eqs. (\ref{eq:Ic_tilde}) and (\ref{eq:p_zero}), the threshold current to move the magnetization from the energetically stable state ($\theta=0$) is 
an oscillating function of $\tau$. 
For example, $I_{\rm c}$ given by Eq. (\ref{eq:Ic_tilde}) becomes $1.9$ mA for $\tau=0.03$ ns, which is smaller than the applied current, $I=2.5$ mA. 
Therefore, the magnetization can move from the initial state, as shown in Fig. \ref{fig:fig4}(a). 
On the other hand, $I_{\rm c}$ becomes $4.4$ mA for $\tau=0.3$ ns, and, therefore, the magnetization stays in the energetically stable state in Fig. \ref{fig:fig4}(b). 
Such a modification of the instability threshold was studied in a vortex oscillator both theoretically and experimentally [\onlinecite{khalsa15,tsunegi16}]. 
For a sufficiently long delay time, the magnetization dynamics becomes highly complex, and Eq. (\ref{eq:Ic_tilde}) does not work. 
The periodic oscillation with the amplitude modulation is found for $\tau=9.3$ ns, 
whereas non-periodic dynamics appears for $\tau=9.6$ ns, as shown in Figs. \ref{fig:fig4}(c) and \ref{fig:fig4}(d). 



Figures \ref{fig:fig4}(e) and \ref{fig:fig4}(f) summarize 
the Lyapunov exponent and bifurcation diagram as a function of the delay time, respectively. 
Note that the magnetization stays in the energetically equilibrium state for $0.3 \le \tau < 1.2$ ns, as in the case shown in Fig. \ref{fig:fig4}(b). 
In such a case, the Lyapunov exponent is negative, indicating that the magnetization saturates to a fixed point. 
On the other hand, chaos appears with increasing the delay time, whereas the periodic oscillations with the amplitude modulation appear for specific values of $\tau$. 
The negative Lyapunov exponent for a short delay time is approximately estimated from a linearized LLG equation [\onlinecite{taniguchi16}] as 
\begin{equation}
  \tilde{\lambda}
  \simeq
  -2\pi \alpha f_{\rm FMR}
  \left(
    1
    -
    \frac{I}{\tilde{I}_{\rm c}}
  \right).
  \label{eq:negative_Ly_theoretical}
\end{equation}
For example, for $\tau=0.3$, Eq. (\ref{eq:negative_Ly_theoretical}) is $-0.09$ GHz, which is close to the numerically estimated value, $-0.11$ GHz. 
We simultaneously emphasize that the limit of $\tau \to 0$ does not correspond to the zero-feedback limit 
(the zero-feedback limit corresponds to $\chi \to 0$). 
Even in the limit of $\tau \to 0$, the feedback current exists and affects the dynamics. 
For example, for $\tau=0.03$, the magnetization shows a limit cycle oscillation, and the Lyapunov exponent is zero. 
Equation (\ref{eq:negative_Ly_theoretical}) works when the magnetization stays at a fixed point, and the delay time $\tau$ is short. 




\section{Theoretical analysis}
\label{sec:Theoretical analysis}

The above numerical results indicate the existence of rich variety of nonlinear dynamics, including chaos, in an STO. 
Although it is difficult to solve the LLG equation exactly due to its nonlinearity, 
let us investigate the physical origin of the complex dynamics with help of an approximated theory, 
which has been known to be useful to analyze nonlinear dynamics 
such as auto-oscillation (limit cycle) [\onlinecite{bertotti09text,taniguchi17}] and synchronization [\onlinecite{taniguchi17PRB}]. 
An auto-oscillation in an STO is excited when the spin-transfer torque balances with the damping torque, and the field torque, $-\gamma \mathbf{m} \times \mathbf{H}$, remains finite. 
The field torque leads to a sustainable oscillation of the magnetization on a constant energy curve 
of the magnetic energy density defined as $E=-M \int d \mathbf{m}\cdot\mathbf{H}$. 
In the present system, the constant energy curve corresponds to the trajectory with a constant zenith angle $\theta=\cos^{-1}m_{z}$, 
where the oscillation frequency, $f(\theta)$, on the constant energy curve is $f(\theta)=\gamma[H_{\rm appl}+(H_{\rm K}-4\pi M)\cos\theta]/(2\pi)$. 
It should be, however, emphasized that there is often an instantaneous imbalance between the spin-transfer torque and damping torque 
because of their different angular dependencies. 
Therefore, strictly speaking, $\theta$ (or $m_{z}$) in the present system is not a constant variable [\onlinecite{taniguchi17}]; see also the inset of Fig. \ref{fig:fig1}(b). 
However, for a sufficiently small damping constant $\alpha$, 
the real trajectory of the auto-oscillation is practically close to a constant energy curve. 
In such a case, it is useful to derive the equation of motion of $\theta$ averaged over the precession period $T(\theta)=1/f(\theta)$ as 
$\overline{d \theta/dt}\equiv (1/T) \oint dt (d\theta/dt)$ 
(see also Appendix \ref{sec:AppendixA} for derivation), 
\begin{equation}
\begin{split}
  \overline{
    \frac{d \theta}{dt}
  }
  =&
  -\alpha 
  \gamma
  \left[
    H_{\rm appl}
    +
    \left(
      H_{\rm K}
      -
      4\pi M
    \right)
    \cos\theta
  \right]
  \sin\theta
\\
  &
  +
  \frac{\gamma H_{\rm s0}}{\lambda \tan\theta}
  \left(
    \frac{1}{\sqrt{1-\lambda^{2}\sin^{2}\theta}}
    -
    1
  \right)
  p(\chi,\tau,\theta),
  \label{eq:LLG_ave}
\end{split}
\end{equation}
where $H_{\rm s0}=\hbar \eta I/(2eMV)$, whereas $p(\chi,\tau,\theta)$ is given by 
\begin{equation}
  p(\chi,\tau,\theta)
  =
  1
  -
  \frac{\chi}{\lambda}
  \cos 2\pi f(\theta)\tau.
  \label{eq:p}
\end{equation}



\begin{figure*}
\centerline{\includegraphics[width=2.0\columnwidth]{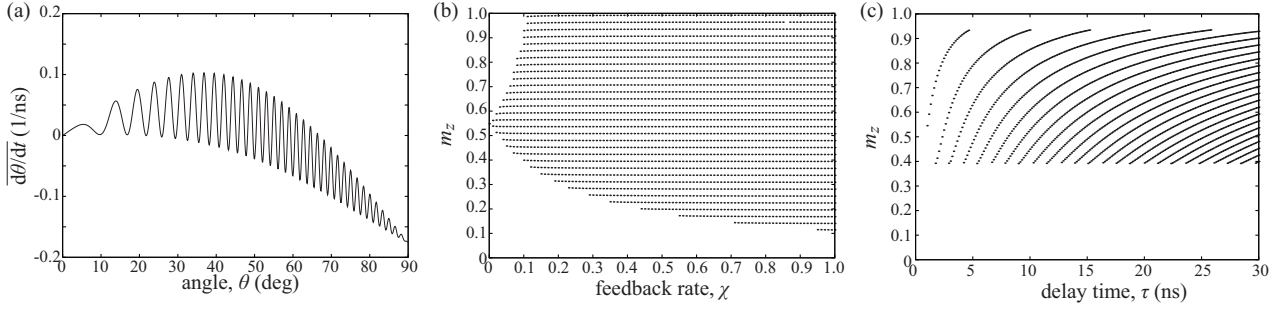}}
\caption{
        (a) The averaged $d \theta/dt$ given by Eq. (\ref{eq:LLG_ave}) solved in the phase space as a function of $\theta=\cos^{-1}m_{z}$. 
            The current, feedback rate, and delay time are $I=2.5$ mA, $\chi=0.10$, and $\tau=30$ ns. 
        (b), (c) Stable fixed points $m_{z}=\cos\theta$ estimated analytically as a function of (b) the feedback rate $\chi$ with $\tau=30$ ns and (c) the delay time $\tau$ with $\chi=0.10$. 
         \vspace{-3ex}}
\label{fig:fig5}
\end{figure*}



The angle $\theta$ satisfying $\overline{d \theta/dt}=0$ and $d (\overline{d \theta/dt})/ d\theta< (>)0$ corresponds to a stable (unstable) fixed point in the reduced phase space [\onlinecite{strogatz01}]. 
In the absence of feedback current, there is only one stable fixed point (attractor), corresponding to auto-oscillation state in real space, in the present STO [\onlinecite{taniguchi17}]. 
On the other hand, Fig. \ref{fig:fig5}(a) shows an example of $\overline{d \theta/dt}$ in the presence of the feedback. 
As shown, several attractors satisfying $\overline{d\theta/dt}=0$ and $d (\overline{d \theta/dt})/ d\theta<0$ appear due to the feedback current.
Figures \ref{fig:fig5}(b) and \ref{fig:fig5}(c) show the attractors $m_{z}=\cos\theta$ as a function of the feedback rate $\chi$ and the delay time $\tau$, respectively. 
It can be understood from these figures that the number of the attractor increases with increasing the feedback rate and/or delay time. 
Let us here call such structures as multiple attractors. 
Although these results are obtained with an approximation mentioned above, 
they are useful to understand the origin of the complex magnetization dynamics found by numerical simulation, as discussed below. 


The multiple attractors originate from the function $p(\chi,\tau,\theta)$ given by Eq. (\ref{eq:p}). 
In the absence of the feedback current ($\chi=0$), the function $p(\chi,\tau,\theta)=1$ is independent of the angle $\theta$. 
On the other hand, in the presence of the feedback current ($\chi \neq 0$), 
several values of the angle $\theta$ give an identical value of $p(\chi,\tau,\theta)$ because the function includes a periodic (cosine) function depending on $\theta$. 
As a result, several $\theta$ can simultaneously satisfy the conditions of the stable fixed point. 


The origin of the complex dynamics found in the numerical simulation is considered to be the existence of multiple attractors. 
Since the attractors locate discretely, as shown in Fig. \ref{fig:fig5}, 
one might consider that once the magnetization is trapped by one of the attractors, it cannot move to the others. 
It should be, however, reminded that the assumption of a constant angle $\theta$ was used in the derivation of Eq. (\ref{eq:LLG_ave}). 
As emphasized above, the real angle $\theta=\cos^{-1}m_{z}$ in a limit cycle slightly oscillates around the fixed point estimated analytically by Eq. (\ref{eq:LLG_ave}) 
because of the instantaneous imbalance between the spin-transfer torque and damping torque. 
As a result, the magnetization can move from one attractor to the other 
when the distance between the attractors is smaller than the oscillation amplitude of the angle $\theta$. 
The transition between the attractors causes the highly complex dynamics shown in Fig. \ref{fig:fig2}, contrary to the system without feedback in which an auto-oscillation state is uniquely determined. 


It is considered that the above analytical theory can be applied to any type of STO, 
although Eq. (\ref{eq:LLG_ave}) was derived for its specific type. 
For example, the complex dynamics 
found in an in-plane magnetized STO [\onlinecite{williame19}] may be caused 
by the same mechanism, i.e., the appearance of multiple attractors due to the existence of feedback current. 
The periodicity of the multiple attractors in this type of STO is described by elliptic functions 
in contrast with Eq. (\ref{eq:p}) where the periodicity is described by a simple trigonometric function; see Appendix \ref{sec:AppendixB}. 



\subsection{Conclusion}
\label{sec:Conclusion}

In conclusion, the nonlinear magnetization dynamics in a spin-torque oscillator was studied by taking into account the effect of spin-transfer torque excited by the feedback current. 
The numerical simulation reveals rich variety of the nonlinear magnetization dynamics, which can be controlled by the feedback parameter. 
The positive Lyapunov exponent for a certain range of the feedback rate indicated the existence of chaos in the spin-torque oscillator, 
whereas transient behavior from the chaotic to the steady state was also observed in another range of the feedback parameter. 
The analytical theory based on the averaged equation of motion revealed that the feedback current results in the multiple attractors in the phase space. 
The number of the attractors increased with increasing the feedback rate and/or delay time. 
An instantaneous imbalance between the spin-transfer torque and damping torque caused a transition between the attractors, and induces the complex magnetization dynamics. 


\section*{Acknowledgement}

The authors are thankful to Joo-Von Kim, Takehiko Yorozu, Sumito Tsunegi, and Shinji Miwa for valuable discussion. 
T. T. is grateful to Satoshi Iba, Aurelie Spiesser, Hiroki Maehara, and Ai Emura for their support and encouragement. 
The results were partially obtained from a project (Innovative AI Chips and Next-Generation Computing Technology Development/(2) 
Development of next-generation computing technologies/Exploration of Neuromorphic Dynamics towards Future Symbiotic Society) commissioned by NEDO. 
K. N. is supported by 
JSPS KAKENHI Grant Numbers JP18H05472, and JP16KT0019.
H. N. is supported by JSPS KAKENHI Grant Number JP18H01135, and JST PRESTO Grant Number JPMJPR16EA.
M. K. is supported by JSPS KAKENHI Grant Numbers JP16H02155, JP17H02857.


\appendix



\section{Averaged LLG equation of perpendicularly magnetized STO}
\label{sec:AppendixA}

Introducing the zenith and azimuth angles $(\theta,\varphi)$ as $\mathbf{m}=(\sin\theta\cos\varphi,\sin\theta\sin\varphi,\cos\theta)$, 
the LLG equation (\ref{eq:LLG}), for $\theta$ is given by 
\begin{equation}
\begin{split}
  \frac{d \theta}{dt}
  =&
  -\frac{\gamma \hbar \eta I[1+\chi\mathbf{m}(t-\tau)\cdot\mathbf{p}]}{2e(1+\lambda \sin\theta \cos\varphi)MV}
  \cos\theta
  \cos\varphi
\\
  &-
  \alpha
  \gamma
  \left[
    H_{\rm appl}
    +
    \left(
      H_{\rm K}
      -
      4\pi M
    \right)
    \cos\theta
  \right]
  \sin\theta,
  \label{eq:LLG_theta}
\end{split}
\end{equation}
where the higher order terms of $\alpha$ are neglected. 
As mentioned in the main text, an auto-oscillation is excited with a trajectory depicting practically on a constant energy curve of 
$E=-M \int d \mathbf{m}\cdot\mathbf{H}=-MH_{\rm appl}\cos\theta-[M(H_{\rm K}-4\pi M)/2]\cos^{2}\theta$. 
The dynamical trajectory on the constant energy curve, 
which is the solution of $d \mathbf{m}/dt=-\gamma \mathbf{m} \times \mathbf{H}$, is given by 
$m_{x}=\sin\theta\cos\omega(\theta)t$, $m_{y}=\sin\theta\sin\omega(\theta)t$, and $m_{z}=\cos\theta$, 
where $\theta$ is constant whereas 
\begin{equation}
  \omega(\theta)
  =
  \gamma
  \left[
    H_{\rm appl}
    +
    \left(
      H_{\rm K}
      -
      4\pi M
    \right)
    \cos\theta
  \right].
\end{equation}
The frequency and period of the auto-oscillation are $f(\theta)=\omega(\theta)/(2\pi)$ and $T(\theta)=1/f(\theta)$, respectively. 
Substituting these solutions, $m_{x}$, $m_{y}$, and $m_{z}$, into Eq. (\ref{eq:LLG_theta}), we find that 
\begin{equation}
\begin{split}
  &
  \frac{1}{T(\theta)}
  \oint 
  dt 
  \frac{d \theta}{dt}
\\
  &=
  -\frac{\gamma \hbar \eta I}{2eMV T(\theta)}
  \int_{0}^{T(\theta)}
  dt 
  \frac{[1 + \chi \sin\theta \cos \omega(t-\tau)] \cos\theta \cos \omega t}{1+\lambda \sin\theta \cos \omega t}
\\
  &-
  \frac{\alpha \gamma}{T(\theta)}
  \int_{0}^{T(\theta)}
  dt 
  \left[
    H_{\rm appl}
    +
    \left(
      H_{\rm K}
      -
      4\pi M 
    \right)
  \right]
  \sin\theta. 
\end{split}
\end{equation}
Using the integral formulas, we find that 
\begin{equation}
\begin{split}
  \frac{1}{T(\theta)}
  \oint 
  dt
  \frac{d\theta}{dt}
  =&
  \frac{\gamma \hbar \eta I}{2e\lambda MV \tan\theta}
  \left(
    \frac{1}{\sqrt{1-\lambda^{2}\sin^{2}\theta}}
    -
    1
  \right)
  p(\chi,\tau,\theta)
\\
  &
  -
  \alpha 
  \gamma
  \left[
    H_{\rm appl}
    +
    \left(
      H_{\rm K}
      -
      4\pi M
    \right)
    \cos\theta
  \right]
  \sin\theta, 
  \label{eq:LLG_ave_Appendix}
\end{split}
\end{equation}
where  $p(\chi,\tau,\theta)$ is given by Eq.(\ref{eq:p}). 
Equation (\ref{eq:LLG_ave_Appendix}) is identical to Eq. (\ref{eq:LLG_ave}). 
The threshold current given by  Eq. (\ref{eq:Ic_tilde}) is the current satisfying $\lim_{\theta \to 0}\overline{d\theta/dt}=0$, 
whereas Eq. (\ref{eq:critical_current}) is Eq. (\ref{eq:Ic_tilde}) in the limit of $\chi \to 0$.


\section{Averaged LLG equation of in-plane magnetized STO}
\label{sec:AppendixB}

In the main text, the multiple attractors are investigated for an STO consisting of a perpendicularly magnetized free layer 
and an in-plane magnetized reference layer. 
On the other hand, previous works had focused on an STO consisting of in-plane magnetized free and reference layers [\onlinecite{li06,yang07,kudo06,williame19}]. 
Therefore, let us show that the in-plane magnetized STO also shows the multiple attractors structure when the spin-transfer torque includes the feedback current. 
In this Appendix, the values of the parameters are derived from Refs. [\onlinecite{taniguchi17PRB,taniguchi18PRB,taniguchi18PRB2}], 
The magnetic field and the strength of the spin-transfer torque of an in-plane magnetized STO are given by 
$\mathbf{H}=H_{\rm K}m_{y}\mathbf{e}_{y}-4\pi M m_{z}\mathbf{e}_{z}$ and $H_{\rm s}=\hbar \eta J/(2eMd)$, respectively, 
where $H_{\rm K}=200$ Oe is an in-plane anisotropy field along the easy ($y$) axis, 
$J$ is the current density, and $d=2.0$ nm is the thickness of the free layer. 
The saturation magnetization and the Gilbert damping constant are $M=1500$ emu/c.c. and $0.01$, respectively. 
The spin polarization $\eta$ is 0.5, whereas the spin-transfer torque asymmetry $\lambda$ is assumed to be zero, for simplicity. 
The spin-polarization direction $\mathbf{p}$ is parallel to the easy axis, $\mathbf{p}=\mathbf{e}_{y}$. 


\subsection{Energy range of in-plane auto-oscillation}

As mentioned in the main text, the averaged LLG equation is derived by assuming an auto-oscillation on a constant energy curve. 
The energy density of an in-plane magnetized ferromagnet is given by 
\begin{equation}
  E
  =
  -\frac{MH_{\rm K}}{2}
  m_{y}^{2}
  +
  \frac{4\pi M^{2}}{2}
  m_{z}^{2}.
  \label{eq:energy}
\end{equation}
The minimum, saddle, and maximum energy densities are $E_{\rm min}=-MH_{\rm K}/2$, $E_{\rm s}=0$, and $E_{\rm max}=4\pi M^{2}/2$, 
corresponding to the magnetization states of $\mathbf{m}=\pm \mathbf{e}_{y}$, $\pm \mathbf{e}_{x}$, and $\pm \mathbf{e}_{z}$, respectively. 
Here, we focus on the auto-oscillation around the easy axis, 
where the corresponding energy density $E$ is in the range of $E_{\rm min} < E < E_{\rm s}$. 
The auto-oscillation is excited when the current density is in the range of $J_{\rm c} < J < J^{*}$ [\onlinecite{taniguchi17PRB,taniguchi18PRB,taniguchi18PRB2}], 
where $J_{\rm c}$ and $J^{*}$ are the critical and switching current densities given by 
\begin{equation}
  J_{\rm c}
  =
  \frac{2 \alpha eMd}{\hbar \eta}
  \left(
    H_{\rm K}
    +
    2\pi M
  \right),
\end{equation}
\begin{equation}
  J^{*}
  =
  \frac{4 \alpha eMd}{\pi \hbar \eta}
  \sqrt{
    4\pi M
    \left(
      H_{\rm K}
      +
      4\pi M
    \right)
  }.
\end{equation}


\subsection{Averaged LLG equation in the absence of feedback current}

The LLG equation averaged over the constant energy curve of $E$ in the in-plane magnetized ferromagnet 
without the feedback current is given by [\onlinecite{taniguchi17PRB}] 
\begin{equation}
  \oint 
  dt 
  \frac{d E}{dt}
  =
  \mathscr{W}_{\rm s}
  +
  \mathscr{W}_{\alpha},
\end{equation}
where $\mathscr{W}_{\rm s}$ and $\mathscr{W}_{\alpha}$ are the work done by the spin-transfer torque and the energy dissipation by the damping torque 
during a precession on a constant energy curve, 
\begin{equation}
\begin{split}
  \mathscr{W}_{\rm s}
  &=
  \gamma M 
  \oint 
  dt H_{\rm s}
  \left[
    \mathbf{p}
    \cdot
    \mathbf{H}
    -
    \left(
      \mathbf{m}
      \cdot
      \mathbf{p}
    \right)
    \left(
      \mathbf{m}
      \cdot
      \mathbf{H}
    \right)
  \right]
\\
  &=
  2\pi M H_{\rm s}
  \frac{2E/M+H_{\rm K}}{\sqrt{H_{\rm K}(H_{\rm K}+4\pi M)}},
  \label{eq:W_s}
\end{split}
\end{equation}
\begin{equation}
\begin{split}
  \mathscr{W}_{\alpha}
  &=
  -\alpha \gamma M
  \oint
  dt 
  \left[
    \mathbf{H}^{2}
    -
    \left(
      \mathbf{m}
      \cdot
      \mathbf{H}
    \right)^{2}
  \right]
\\
  &=
  -4\alpha M 
  \sqrt{
    \frac{4\pi M-2E/M}{H_{\rm K}}
  }
  \left[
    \frac{2E}{M}
    \mathsf{K}(k)
    +
    H_{\rm K}
    \mathsf{E}(k)
  \right],
  \label{eq:W_alpha}
\end{split}
\end{equation}
where $\mathsf{K}(k)=\int_{0}^{1}dx/\sqrt{(1-x^{2})(1-k^{2}x^{2})}$ 
and $\mathsf{E}(k)=\int_{0}^{1}dx\sqrt{(1-k^{2}x^{2})/(1-x^{2})}$ are the first and second kind of complete elliptic integral with the modulus $k$: 
\begin{equation}
  k
  =
  \sqrt{
    \frac{4\pi M(H_{\rm K}+2E/M)}{H_{\rm K}(4\pi M-2E/M)}
  }.
\end{equation}
The precession period $T(E)$ on a constant energy curve of $E$ is 
\begin{equation}
  T(E)
  =
  \frac{4 \mathsf{K}(k)}{\gamma \sqrt{H_{\rm K}(4\pi M-2E/M)}}.
\end{equation}



\begin{figure}
\centerline{\includegraphics[width=1.0\columnwidth]{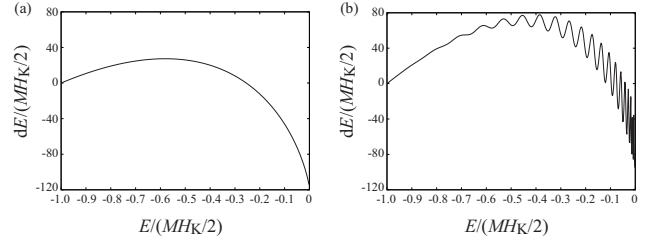}}
\caption{
        The averaged energy change, $dE\equiv \oint dt (dE/dt)$, an in-plane magnetized ferromagnet as a function of the energy density $E$. 
        The vertical and horizontal axes are renormalized by $MH_{\rm K}/2$. 
        The feedback current is (a) zero and (b) $\chi=0.10$ with $\tau=3$ ns. 
         \vspace{-1ex}}
\label{fig:fig6}
\end{figure}



Figure \ref{fig:fig6}(a) shows an example of $dE \equiv \oint dt (dE/dt)$ in the absence of the feedback current, 
where the current density is chosen to be $J=(J_{\rm c}+J^{*})/2$. 
The energy density $E$ satisfying $dE=0$ and $d(dE)/dE<0$ corresponds to a stable attractor. 
As in the case of the STO in the main text, there is only one attractor in this system. 


\subsection{Work done by feedback current}

Now let us consider the role of the feedback current. 
In the presence of the feedback current, the spin-transfer torque performs an additional work given by 
\begin{equation}
  W_{\rm s}^{\chi}
  \equiv 
  \gamma M 
  \oint 
  dt 
  H_{\rm s}
  \chi 
  \mathbf{m}(t-\tau)
  \cdot 
  \mathbf{p}
  \left[
    \mathbf{p}
    \cdot
    \mathbf{H}
    -
    \left(
      \mathbf{m}
      \cdot
      \mathbf{p}
    \right)
    \left(
      \mathbf{m}
      \cdot
      \mathbf{H}
    \right)
  \right],
  \label{eq:Ws_chi_def}
\end{equation}
where we assume that the feedback current density is given by $\chi J \mathbf{m}(t-\tau)\cdot\mathbf{p}$. 
The averaged LLG equation in the presence of the feedback current becomes 
\begin{equation}
  \oint 
  dt 
  \frac{dE}{dt}
  =
  \mathscr{W}_{\rm s}
  +
  \mathscr{W}_{\rm s}^{\chi}
  +
  \mathscr{W}_{\alpha}.
\end{equation}

To evaluate $\mathscr{W}_{\rm s}^{\chi}$, it is useful to note that 
the solution of the magnetization oscillating around the easy axis on a constant energy curve of $E$ is given by [\onlinecite{taniguchi17PRB}] 
\begin{equation}
  m_{x}(t)
  =
  \sqrt{
    1
    +
    \frac{2E}{MH_{\rm K}}
  }
  {\rm sn}
  \left[
    \frac{4 \mathsf{K}(k)}{T(E)}
    t, k
  \right],
\end{equation}
\begin{equation}
  m_{y}(t)
  =
  \sqrt{
    \frac{4\pi M-2E/M}{H_{\rm K}+4\pi M}
  }
  {\rm dn}
  \left[
    \frac{4 \mathsf{K}(k)}{T(E)}
    t, k
  \right],
\end{equation}
\begin{equation}
  m_{z}(t)
  =
  \sqrt{
    \frac{H_{\rm K}+2E/M}{H_{\rm K}+4\pi M}
  }
  {\rm cn}
  \left[
    \frac{4 \mathsf{K}(k)}{T(E)}
    t, k
  \right],
\end{equation}
where ${\rm sn}(u,k)$, ${\rm dn}(u,k)$, and ${\rm cn}(u,k)$ are the Jacobi elliptic functions with $u=4 \mathsf{K}(k)t/T(E)$. 
Introducing a new variable $x={\rm sn}(u,k)$, 
Eq. (\ref{eq:Ws_chi_def}) becomes 
\begin{widetext}
\begin{equation}
  \mathscr{W}_{\rm s}^{\chi}
  =
  \frac{4 \chi M H_{\rm s}}{\sqrt{H_{\rm K}(4\pi M-2E/M)}}
  \int_{0}^{1}
  dx 
  \frac{\left[ H_{\rm K}m_{y}-m_{y}(H_{\rm K}m_{y}^{2}-4\pi M m_{z}^{2}) \right]}{\sqrt{(1-x^{2})(1-k^{2}x^{2})}}
  m_{y}(t-\tau),
  \label{eq:Ws_chi_1}
\end{equation}
\end{widetext}
where ${\rm dn}(u,k)$ and ${\rm cn}(u,k)$ in $m_{y}(t)$ and $m_{z}(t)$ are replaced by $\sqrt{1-k^{2}x^{2}}$ and $\sqrt{1-x^{2}}$, respectively. 
On the other hand, $m_{y}(t-\tau)$ in Eq. (\ref{eq:Ws_chi_1}) is given by [\onlinecite{byrd71}]
\begin{widetext}
\begin{equation}
\begin{split}
  m_{y}(t-\tau)
  &=
  \sqrt{
    \frac{4\pi M-2E/M}{H_{\rm K}+4\pi M}
  }
  \frac{{\rm dn}(u,k) {\rm dn}(v,k) + k^{2} {\rm sn}(u,k) {\rm sn}(v,k) {\rm cn}(u,k) {\rm cn}(v,k)}{1-k^{2}{\rm sn}^{2}(u,k) {\rm sn}^{2}(v,k)}
\\
  &=
  \sqrt{
    \frac{4\pi M-2E/M}{H_{\rm K}+4\pi M}
  }
  \frac{{\rm dn}(v,k) \sqrt{1-k^{2}x^{2}} + k^{2} {\rm sn}(v,k) {\rm cn}(v,k) x \sqrt{1-x^{2}}}{1-k^{2}{\rm sn}^{2}(v,k) x^{2}},
  \label{eq:my_delay}
\end{split}
\end{equation}
\end{widetext}
where $v= 4 \mathsf{K}(k) \tau/T(E)$. 
Equation (\ref{eq:my_delay}) indicates that the multiple attractors originate from the periodicity of the elliptic function. 
In contrast with Eqs. (\ref{eq:W_s}) and (\ref{eq:W_alpha}), the analytical expression of Eq. (\ref{eq:Ws_chi_1}) is complex; see next section. 
Therefore, we evaluate Eq. (\ref{eq:Ws_chi_1}) numerically. 

Figure \ref{fig:fig6}(b) shows $\oint dt (dE/dt)$ in the presence of the feedback current, 
where $\chi=0.10$ and $\tau=3$ ns. 
As shown, the multiple attractors appear, as in the STO studied in the main text. 
Therefore, we consider that the chaotic dynamics studied in Ref. [\onlinecite{williame19}] might be also related to the multiple attractors. 


\subsection{Analytical expression of $\mathscr{W}_{\rm s}^{\chi}$}

Substituting Eq. (\ref{eq:my_delay}) into Eq. (\ref{eq:Ws_chi_1}),  $\mathscr{W}_{\rm s}^{\chi}$ is rewritten as 
\begin{equation}
\begin{split}
  \mathscr{W}_{\rm s}^{\chi}
  &=
  \frac{4 \chi M H_{\rm s}}{\sqrt{H_{\rm K}(4\pi M-2E/M)}}
  \sum_{\ell=1}^{5}
  I_{\ell},
\end{split}
\end{equation}
where we introduce $I_{\ell}$ as 
\begin{equation}
\begin{split}
  I_{1}
  &=
  c_{y}^{2}
  H_{\rm K}
  {\rm dn}(v,k)
  \int_{0}^{1}
  dx 
  \frac{\sqrt{1-k^{2}x^{2}}}{\sqrt{1-x^{2}}[1-k^{2}{\rm sn}^{2}(v,k)x^{2}]}
\\
  &
  \equiv 
  c_{y}^{2}
  H_{\rm K}
  {\rm dn}(v,k)
  \tilde{I}_{1},
\end{split}
\end{equation}
\begin{equation}
\begin{split}
  I_{2}
  &=
  -c_{y}^{4}
  H_{\rm K}
  {\rm dn}(v,k)
  \int_{0}^{1}
  dx 
  \frac{(1-k^{2}x^{2})^{3/2}}{\sqrt{1-x^{2}} [1-k^{2}{\rm sn}^{2}(v,k)x^{2}]}
\\
  &
  \equiv
  -c_{y}^{4}
  H_{\rm K}
  {\rm dn}(v,k)
  \tilde{I}_{2},
\end{split}
\end{equation}
\begin{equation}
\begin{split}
  I_{3}
  &=
  c_{y}^{2}
  c_{z}^{2}
  4\pi M 
  {\rm dn}(v,k)
  \int_{0}^{1}
  dx 
  \frac{\sqrt{(1-x^{2})(1-k^{2}x^{2})}}{1-k^{2}{\rm sn}^{2}(v,k)x^{2}}
\\
  &
  \equiv
  c_{y}^{2}
  c_{z}^{2}
  4\pi M 
  {\rm dn}(v,k)
  \tilde{I}_{3},
\end{split}
\end{equation}
\begin{widetext}
\begin{equation}
\begin{split}
  I_{4}
  &=
  c_{y}^{2}
  \left[
    \left(
      1
      -
      c_{y}^{2}
    \right)
    H_{\rm K}
    +
    c_{z}^{2}
    4\pi M
  \right]
  k^{2}
  {\rm sn}(v,k)
  {\rm cn}(v,k)
  \int_{0}^{1}
  dx 
  \frac{x}{1-k^{2}{\rm sn}^{2}(v,k)x^{2}}
\\
  &\equiv
  c_{y}^{2}
  \left[
    \left(
      1
      -
      c_{y}^{2}
    \right)
    H_{\rm K}
    +
    c_{z}^{2}
    4\pi M
  \right]
  k^{2}
  {\rm sn}(v,k)
  {\rm cn}(v,k)
  \tilde{I}_{4},
\end{split}
\end{equation}
\begin{equation}
\begin{split}
  I_{5}
  &=
  c_{y}^{2}
  \left(
    c_{y}^{2}
    H_{\rm K}
    k^{2}
    -
    c_{z}^{2}
    4\pi M
  \right)
  k^{2}
  {\rm sn}(v,k)
  {\rm cn}(v,k)
  \int_{0}^{1}
  dx 
  \frac{x^{3}}{1-k^{2}{\rm sn}^{2}(v,k)x^{2}}
\\
  &\equiv
  c_{y}^{2}
  \left(
    c_{y}^{2}
    H_{\rm K}
    k^{2}
    -
    c_{z}^{2}
    4\pi M
  \right)
  k^{2}
  {\rm sn}(v,k)
  {\rm cn}(v,k)
  \tilde{I}_{5}.
\end{split}
\end{equation}
\end{widetext}
Here, we introduce the following notations, for simplicity.
\begin{align}
  c_{y}
  =
  \sqrt{
    \frac{4\pi M-2E/M}{H_{\rm K}+4\pi M}
  },
&&
  c_{z}
  =
  \sqrt{
    \frac{H_{\rm K}+2E/M}{H_{\rm K}+4\pi M}
  }.
\end{align}

The integrals $\tilde{I}_{\ell}$ ($\ell=1-5$) can be performed as 
\begin{widetext}
\begin{equation}
\begin{split}
  \tilde{I}_{1}
  &=
  \int_{0}^{1}
  dx 
  \frac{\sqrt{1-k^{2}x^{2}}}{\sqrt{1-x^{2}}[1-k^{2}{\rm sn}^{2}(v,k)x^{2}]}
\\
  &=
  \frac{1}{{\rm sn}^{2}(v,k)}
  \int_{0}^{1}
  \frac{dx}{\sqrt{(1-x^{2})(1-k^{2}x^{2})}}
  -
  \frac{1-{\rm sn}^{2}(v,k)}{{\rm sn}^{2}(v,k)}
  \int_{0}^{1}
  \frac{dx}{\sqrt{(1-x^{2})(1-k^{2}x^{2})} [1-k^{2}{\rm sn}^{2}(v,k)x^{2}]}
\\
  &=
  \frac{\mathsf{K}(k)}{{\rm sn}^{2}(v,k)}
  -
  \frac{{\rm cn}^{2}(v,k)}{{\rm sn}^{2}(v,k)}
  \Pi[k^{2}{\rm sn}^{2}(v,k),k],
\end{split}
\end{equation}
\begin{equation}
\begin{split}
  \tilde{I}_{2}
  &=
  \int_{0}^{1} 
  dx 
  \frac{(1-k^{2}x^{2})^{3/2}}{\sqrt{1-x^{2}} [1-k^{2}{\rm sn}^{2}(v,k)x^{2}]}
\\
  &=
  \frac{1}{{\rm sn}^{2}(v,k)}
  \int_{0}^{1}
  dx
  \sqrt{
    \frac{1-k^{2}x^{2}}{1-x^{2}}
  }
  -
  \frac{{\rm cn}^{2}(v,k)}{{\rm sn}^{2}(v,k)}
  \int_{0}^{1}
  dx 
  \frac{\sqrt{1-k^{2}x^{2}}}{\sqrt{1-x^{2}}[1-k^{2}{\rm sn}^{2}(v,k)x^{2}]}
\\
  &=
  \frac{\mathsf{E}(k)}{{\rm sn}^{2}(v,k)}
  -
  \frac{{\rm cn}^{2}(v,k)}{{\rm sn}^{2}(v,k)}
  \tilde{I}_{1},
\end{split}
\end{equation}
\begin{equation}
\begin{split}
  \tilde{I}_{3}
  &=
  \int_{0}^{1}
  dx 
  \frac{\sqrt{(1-x^{2})(1-k^{2}x^{2})}}{1-k^{2}{\rm sn}^{2}(v,k)x^{2}}
\\
  &=
  \frac{1}{k^{2}{\rm sn}^{2}(v,k)}
  \int_{0}^{1}
  dx
  \sqrt{
    \frac{1-k^{2}x^{2}}{1-x^{2}}
  }
  -
  \frac{{\rm dn}^{2}(v,k)}{k^{2}{\rm sn}^{2}(v,k)}
  \int_{0}^{1}
  dx
  \frac{\sqrt{1-k^{2}x^{2}}}{\sqrt{1-x^{2}} [1-k^{2}{\rm sn}^{2}(v,k)x^{2}]}
\\
  &=
  \frac{\mathsf{E}(k)}{k^{2} {\rm sn}^{2}(v,k)}
  -
  \frac{{\rm dn}^{2}(v,k)}{k^{2} {\rm sn}^{2}(v,k)}
  \tilde{I}_{1},
\end{split}
\end{equation}
\end{widetext}
\begin{equation}
\begin{split}
  \tilde{I}_{4}
  &=
  \int_{0}^{1}
  dx
  \frac{x}{1-k^{2}{\rm sn}^{2}(v,k)x^{2}}
\\
  &=
  -\frac{\log[1-k^{2}{\rm sn}^{2}(v,k)]}{2k^{2}{\rm sn}^{2}(v,k)}
\\
  &=
  -\frac{\log {\rm dn}(v,k)}{k^{2}{\rm sn}^{2}(v,k)},
\end{split}
\end{equation}
\begin{widetext}
\begin{equation}
\begin{split}
  \tilde{I}_{5}
  &=
  \int_{0}^{1}
  dx 
  \frac{x^{3}}{1-k^{2}{\rm sn}^{2}(v,k)x^{2}}
\\
  &=
  -\frac{1}{k^{2}{\rm sn}^{2}(v,k)}
  \int_{0}^{1}
  dx x
  +
  \frac{1}{k^{2}{\rm sn}^{2}(v,k)}
  \int_{0}^{1}
  dx 
  \frac{x}{1-k^{2}{\rm sn}^{2}(v,k)x^{2}}
\\
  &=
  -\frac{1}{2k^{2}{\rm sn}^{2}(v,k)}
  +
  \frac{\tilde{I}_{4}}{k^{2}{\rm sn}^{2}(v,k)},
\end{split}
\end{equation}
\end{widetext}
where $\Pi(a^{2},k)=\int_{0}^{1}dx/[(1-a^{2}x^{2})\sqrt{(1-x^{2})(1-k^{2}x^{2})}]$ is the third kind of complete elliptic integral. 





\end{document}